\documentclass{article}
\def\1ad{\mbox{\normalsize $^1$}}
\def\2ad{\mbox{\normalsize $^2$}}
\def\3ad{\mbox{\normalsize $^3$}}
\def\4ad{\mbox{\normalsize $^4$}}
\def\5ad{\mbox{\normalsize $^5$}}
\def\6ad{\mbox{\normalsize $^6$}}
\def\7ad{\mbox{\normalsize $^7$}}
\def\8ad{\mbox{\normalsize $^8$}}
\def\makefront{
    \vspace*{1cm}\begin{center}
    \def\sp{
	\renewcommand{\thefootnote}{\fnsymbol{footnote}}
	\footnote[1]{corresponding author~~E-mail: \email_speaker}
	\renewcommand{\thefootnote}{\arabic{footnote}}
    }
    \def\newtitleline{\\ \vskip 5pt}
    {\Large\bf\titleline}\\
    \vskip 1truecm
    {\large\bf\authors}\\
    \vskip 5truemm
    \addresses
    \end{center}
    \vskip 1truecm
    {\bf Abstract:}
    \abstracttext
    \vskip 1truecm
}

\def\a{\alpha}
\def\b{\beta}
\def\g{\gamma}

\def\eps{\epsilon}
\def\ve{\varepsilon}
\def\z{\zeta}

\def\th{\theta}
\def\vt{\vartheta}

\def\la{\lambda}
\def\m{\mu}
\def\n{\nu}
\def\r{\rho}
\def\s{\sigma}
\def\p{\phi}

\def\x{\xi}

\def\Ps{\Psi}

\newcommand{\C}{\mathbb C}
\newcommand{\R}{\mathbb R}
\newcommand{\Z}{\mathbb Z}
\newcommand{\NN}{\mathbb N}
\newcommand{\Hcal}{{\cal H}}

\newcommand{\Tcal}{{\cal T}}
\def\>{\rangle}
\def\<{\langle}
\def\={\ =\ }
\def\+{\dagger}
\def\e{\textrm{e}}
\def\ii{\textrm{i}}
\def\N2{$N\,{=}\,2$}
\def\pa{\partial}
\def\diff{\textrm{d}}
\def\tr{{\textrm tr}}
\def\sfrac#1#2{{\textstyle\frac{#1}{#2}}}
\def\rd#1{\buildrel{_{_{\hskip 0.01in}\rightarrow}}\over{#1}}
\def\ld#1{\buildrel{_{_{\hskip 0.01in}\leftarrow}}\over{#1}}

\def\com{\buildrel{\th\to0}\over{\longrightarrow}}
\newcommand{\bsy}[1]{\boldsymbol{#1}}

\newcommand{\yb}{{\bar y}}
\newcommand{\zb}{{\bar z}}

\newcommand{\lb}{\bar{\la}}
\newcommand{\mb}{\bar{\mu}}
\newcommand{\adag}{a^{\dagger}}
\newcommand{\bdag}{b^{\dagger}}
\newcommand{\cdag}{c^{\dagger}}
\newcommand{\gt}{\tilde{g}}
\newcommand{\Tdag}{T^{\dagger}}
\newcommand{\Th}{\widehat{T}}
\newcommand{\Tch}{\widehat{\Tcal}}
\newcommand{\gh}{\widehat{\gamma}}

\def\be{\begin{equation}}
\def\ee{\end{equation}}
\def\bea{\begin{eqnarray}}
\def\eea{\end{eqnarray}}
\usepackage{amssymb}
\usepackage{amsmath}
\usepackage{amscd}
\usepackage{latexsym}
\usepackage{ifthen}
\begin {document}                 

\begin{flushright}      
hep-th/0401158\\        
ITP--UH--35/03\\        
\end{flushright}        

\def\titleline{
Noncommutative Instantons and Solitons \footnote{
Talk presented at the 27th Johns Hopkins Workshop in G\"oteborg
and at the 36th International Symposium Ahrenshoop in Berlin, 
both in August 2003.}
}
\def\email_speaker{
{\tt 
lechtenf@itp.uni-hannover.de
}}
\def\authors{
Olaf Lechtenfeld
}
\def\addresses{
Institut f\"ur Theoretische Physik, Universit\"at Hannover\\
Appelstra\ss{}e 2, D--30167 Hannover, Germany\\
Email: lechtenf@itp.uni-hannover.de
}
\def\abstracttext{
I explain how to construct noncommutative BPS configurations
in four and lower dimensions by solving linear matrix equations.
Examples are instantons in $D{=}4$ Yang-Mills, monopoles in
$D{=}3$ Yang-Mills-Higgs, and (moving) solitons in $D{=}2{+}1$
Yang-Mills-Higgs. Some emphasis is on the latter as a showcase
for the dressing method.
}
\large
\makefront


\section{Self-duality and BPS equations}
\noindent
In this talk I shall present a powerful method for and results of 
constructing classical field configurations with finite action or energy 
in four-dimensional noncommutative gauge theory and its lower-dimensional
descendants:
\bea
D=4{+}0\quad \textrm{instantons} \qquad&\longrightarrow&\qquad
D=3{+}0\quad \textrm{monopoles} \\
D=2{+}2\quad \textrm{``waves''} \quad\qquad&\longrightarrow&\qquad
D=2{+}1\quad \textrm{solitons}
\eea
I am setting up the formalism in such a way that it is completely 
transparent to the (Moyal-type) noncommutative deformation. In other words, 
the noncommutative equations below differ from the commutative ones merely
in the interpretation of the symbols or their product (stars are suppressed).
This will be briefly explained in Section~6.
The Yang-Mills field equations are implied by first-order 
(self-duality or BPS) equations:
\bea
D{=}4: && D^\mu F_{\mu\nu}=0 \quad\Longleftarrow\quad
F_{\mu\nu}=\sfrac12\ve_{\mu\nu\rho\lambda}F^{\rho\lambda}\\
\downarrow \ \quad && \qquad 
\downarrow \qquad\qquad\qquad\qquad\qquad \downarrow \nonumber\\
D{=}3: && \begin{smallmatrix} D^a F_{ab}\=\p D_b \p\\
                                      D^a D_a \p \=0 \end{smallmatrix} 
\quad\Longleftarrow\quad F_{ab}=\ve_{abc}D^c\p
\eea
where $F$ and $\p$ are $u(n)$ valued and Greek indices run from 1 to 4 
while Latin ones stop at~3. In complex coordinates (note the signs!)
$y=x^1+\ii x^2$ and $z=x^3\mp\ii x^4$
the self-duality equation $F=*F$ becomes
\be \label{selfdual}
[D_y,D_z]\=0\=[D_\yb,D_\zb] \qquad\textrm{and}\qquad 
[D_y,D_\yb]\pm[D_z,D_\zb]\=0
\ee
where the upper and lower signs belong to the signatures (4,0)
and (2,2), respectively.
Dimensional reduction to $D{=}3$ is accomplished via
\be
\pa_4=0 \ ,\quad A_4=\p 
\quad\textrm{for}\quad D=3{+}0 \qquad\textrm{or}\qquad
\pa_3=0 \ ,\quad A_3=\p
\quad\textrm{for}\quad D=2{+}1
\quad.
\ee

\section{Lax pair}
\noindent
The three self-duality equations~(\ref{selfdual})
are the compatibility conditions of the linear system
\be \label{linsys}
( D_\yb -  \la D_z )\,\Ps(x,\la)\ =\ 0\ =\
( D_\zb \pm\la D_y )\,\Ps(x,\la) \quad,
\ee
where $\Ps(x,\la) \in\textrm{U}(n)$ is a matrix function holomorphic in 
the spectral parameter $\la\in\C P^1\simeq S^2$.
{}From the auxiliary function $\Ps$ the gauge potential can be recovered via
\be \label{AfromPsi}
A_\yb - \la A_z \= \Ps (\pa_\yb - \la\pa_z) \Ps^{-1} 
\qquad\textrm{and}\qquad 
A_\zb\pm\la A_y \= \Ps (\pa_\zb\pm\la\pa_y) \Ps^{-1} \quad.
\ee 
In addition, antihermiticity of $A$ corresponds to a normalization
condition for~$\Ps$,
\be \label{norm}
A_\mu^\+ \= -A_\mu \qquad\Longleftrightarrow\qquad
\Ps(x,\la)\,\Ps(x,\mp 1/\lb)^\+ \= \bsy{1} \quad,
\ee
which involves a reflection of $\la$ on the unit circle.

\section{Gauge fixing}
\noindent
Out of the three self-duality equations~(\ref{selfdual}),
the (2,0) part $\ F_{yz}=0\ $ and the (0,2) part $\ F_{\yb\zb}=0\ $
are solved by
\be 
\begin{matrix}
A_y\=g^{-1}\pa_y\,g \\[6pt] A_z\=g^{-1}\pa_z\,g
\end{matrix}
\qquad\textrm{and}\qquad
\begin{matrix}
A_{\yb}\=\gt^{-1}\pa_{\yb}\,\gt \\[6pt] A_{\zb}\=\gt^{-1}\pa_{\zb}\,\gt
\end{matrix}
\ee
for ${g}, {\gt} \in \textrm{GL}(n,\C)$, 
possibly with the restriction that
\be
A_y^\+ = -A_\yb \quad\textrm{and}\quad A_z^\+ = -A_\zb
\qquad\Longrightarrow\qquad
\gt = (g^\+)^{-1} \quad.
\ee
Allowing for $\Ps\in\textrm{GL}(n,\C)$
I may transform to the so-called hermitean gauge,
\be
\Ps \to \gt\Ps\ ,\quad
A_\yb \to 0\ ,\quad A_\zb \to 0\ ,\quad 
A_y \to {h}^{-1}\pa_y\,{h}\ ,\quad A_z \to {h}^{-1}\pa_z\,{h} \quad,
\ee
where $\ h=g\gt^{-1}=gg^\+=h^\+$. 
This gauge eliminates half of~$A$, but the price to pay is that now
$\ A_y^\+ \neq -A_\yb\ $ and $\ A_z^\+ \neq -A_\zb\ $ as well as
\be
\Ps(x,\la)\,\Ps(x,\mp 1/\lb)^\+\={\gt g^{-1}}\={h^{-1}}\ \neq\ \bsy{1} \quad.
\ee
The remaining (1,1) part of the self-duality equations~(\ref{selfdual}) 
produces a second-order equation for the prepotential $h$:
\be 
F_{y\yb}\pm F_{z\zb}\=0 \qquad\Longrightarrow\qquad
\pa_\yb (h^{-1}\pa_y h) \pm \pa_\zb (h^{-1}\pa_z h) \=0 \quad.
\ee
In the hermitean gauge the linear system~(\ref{linsys}) reads
\be 
(\pa_\yb - \la\pa_z)\Ps\=   \la A_z\Ps \qquad\textrm{and}\qquad
(\pa_\zb\pm\la\pa_y)\Ps\=\mp\la A_y\Ps \quad.
\ee
{}From its solution it is, in principle, always possible to retrieve
an antihermitean gauge potential by an appropriate gauge transformation.

\section{Dressing method}
\noindent
In hermitean gauge, the reconstruction (\ref{AfromPsi}) of $A$ from $\Ps$ 
simplifies to
\be \label{res1}
A_z \=\Ps(\la)\,(\pa_z - \sfrac{1}{\la}\pa_\yb)\,\Ps(\la)^{-1}
\qquad\textrm{and}\qquad
A_y \=\Ps(\la)\,(\pa_y\pm\sfrac{1}{\la}\pa_\zb)\,\Ps(\la)^{-1}
\ee
where $\Ps$ is subject to 
\be \label{res2}
h^{-1} \= \Ps(\la)\,\Ps(\mp 1/\lb)^\+ \quad.
\ee
Since $\la\in\C P^1$ a nonconstant matrix function $\Ps(\la)$ 
cannot be globally holomorphic.
Hence, it must have poles at $\la{=}\m_k$, $\ k{=}1,\ldots,m$.
The power of holomorphy then enables us to find $\Ps$ without knowing $A$,
just by fixing its pole structure!

The dressing method~\cite{dressing} builds up $\Ps(x,\la)$ multiplicatively:
$\Ps_{k\textrm{\ poles}} = \chi_k\cdot\Ps_{k-1\textrm{\ poles}}$,
employing the ansatz 
\be
\chi_k(x,\la) \= \bsy{1}-
\frac{\la(1{\pm}\m_k\bar{\m}_k)}{\la-\m_k}\,P_k(x)
\ee
with moduli $\m_k$ and matrices $P_k(x)$, and starting from
the trivial seed solution $\Ps_0=\bsy{1}$.
An $m$-fold repetition of this dressing transformation yields
\be 
\Ps_m(x,\la) \= \prod_{k=1}^m
\Bigl( \bsy{1}-\frac{\la(1{\pm}\m_k\bar{\m}_k)}{\la-\m_k}\,P_k(x) \Bigr) \=
\bsy{1} - \sum_{k=1}^m \frac{\la R_k(x)}{\la-\m_k}
\ee
if all moduli $\m_k$ are mutually different.

\section{Single-pole ansatz}
\noindent
A lot can be learned already from the simplest situation,
namely $m{=}1$ (a single pole and moduli $\mu$):
\be 
\Ps(x,\la)\=\bsy{1}-\frac{\la(1{\pm}\m\mb)}{\la-\m}\,P(x) 
\ee
where the group-valued but $\la$-independent function $P$ is to be determined.
It is crucial to observe that the left hand sides of (\ref{res1}) and 
(\ref{res2}) are $\la$-independent, implying that their right hand sides must
have vanishing residues for the poles at $\la{=}\m$ and $\la{=}{\mp}1/\mb$.
A short computation reveals the following:
\bea
(\ref{res2})\qquad &\Longrightarrow &\qquad 
    P^2 \= P \= P^\+ 
    \qquad\textrm{hermitean projector} \\[8pt]
    &\Longleftrightarrow &\qquad 
    P\=T\,\sfrac{1}{\Tdag T}\,\Tdag 
    \qquad\textrm{$n\times r$(ank) matrix $T(x)$} \phantom{X}
\eea
\bea
(\ref{res1})\qquad &\Longrightarrow &\qquad 
    P\,(\pa_\yb{-}\m\pa_z)\,P \= 0 \=
    (\bsy{1}{-}P)\,(\pa_z{\pm}\mb\pa_\yb)\,P  \nonumber\\[8pt]
    &&\qquad 
    P\,(\pa_\zb{\pm}\m\pa_y)\,P \= 0 \=
    (\bsy{1}{-}P)\,(\pa_y{-}\mb\pa_\zb)\,P  \nonumber\\[8pt]
    &\Longleftrightarrow &\qquad
    (\bsy{1}{-}P)\,L\,T\=0  \qquad\textrm{with}\quad
    L := \begin{cases}
    \pa_z{\pm}\mb\pa_\yb \\ \pa_y{-}\mb\pa_\zb \end{cases} \\[8pt]
    &\Longleftrightarrow &\qquad
    L\,T\= T\,\g 
    \qquad\textrm{for some $r\times r$ matrix $\g$} \quad. \label{eigen}
\eea
I conclude: Every collection $T(x)$ of $r$ simultaneous ``eigenvectors'' 
of the differential operators~$L$ gives rise to a valid projector~$P(x)$
which, in turn, yields a prepotential and a self-dual gauge connection:
\be
h^{-1}\=\bsy{1}-(1{\pm}\m\mb)\,P \qquad\textrm{and}\qquad 
A_z\=\sfrac{1{\pm}\m\mb}{\m}\,\pa_\yb P \quad,\qquad
A_y\=\mp\sfrac{1{\pm}\m\mb}{\m}\,\pa_\zb P \quad.
\ee

\section{Noncommutative deformation}
\noindent
Up to now it seems that I have just reformulated rather old results. 
However, everything still makes sense if I understand all products of
functions in the deformed sense, i.e.
\bea
(f\ g)(x) \qquad\textrm{means}\qquad
(f\star g)(x)&\=&f(x)\,\exp\,\bigl\{ \sfrac{\ii}{2}
{\ld{\partial}}_\m \,\th^{\m\n}\, {\rd{\partial}}_\n \bigr\}\,g(x)\\[4pt]
&\=&f(x)\,g(x)+\sfrac{\ii}{2}\th^{\m\n}\,(\pa_{\m}f)(x)\,(\pa_{\n}g)(x)+\ldots
\nonumber\eea
with $\ \th^{\m\n}=-\th^{\n\m}=\textrm{constant}\ $ in $D=4{+}0$ dimensions.
The coordinate functions then obey the (star) commutation rule
\be
x^\m\star x^\n-x^\n\star x^\m\ =\ \ii\th^{\m\n} \quad.
\ee
For a given noncommutativity matrix $(\th^{\mu\nu})$ I can choose an 
orthonormal basis in which 
\be
\bigl(\th^{\m\n}\bigr) \= \biggl( \begin{smallmatrix}
0 & \th & 0 & 0 \\ -\th & 0 & 0 & 0 \\ 0 & 0 & 0 & \th'\\ 0 & 0 & -\th'& 0
\end{smallmatrix} \biggr) \quad.
\ee
In this talk I specialize to $\ \th'=\th\ $ in $4{+}0$ dimensions 
(self-dual noncommutativity) while lower dimensions enforce $\ \th'=0$. 
For my choice of complex coordinates this implies that (note the asymmetry!)
\be
y\star\yb-\yb\star y\ =\ 2\th\ =\ \zb\star z-z\star\zb \quad.
\ee

Via the Moyal-Weyl correspondence, this structure (the deformed function 
algebra) can be realized equivalently by an operator algebra with the usual
(compositional) product,
\be
\bigl( f(y,\yb,z,\zb),\star\bigr)\qquad\simeq\qquad
\bigl( F(a,\adag,\bdag,b),\cdot \bigr) \quad.
\ee
The latter is generated by two sets $(a,\adag;b,\bdag)$ of oscillator 
annihilation and creation operators subject to the Heisenberg algebra
\be
[a,\adag] \= 1 \= [b,\bdag]
\ee
which can be represented on a Fock space~$\Hcal_2$.
Putting $2\th{=}1$ for convenience, the Moyal-Weyl map and its inverse
operate as follows,
\be
F\=\textrm{Weyl-order}\ \bigl[f(a,\adag,\bdag, b)\bigr]
\qquad\textrm{and}\qquad f\=F_\star(y,\yb,z,\zb) \quad,
\ee
where star multiplication is implied when writing out the Weyl symbol
$F_\star$ in terms of the coordinates.
It is also worth noting that
\bea &
\pa_y f  \ \simeq\ -[\adag,F] \quad,\quad
\pa_\yb f\ \simeq\  [a,F] \quad,\quad 
\pa_z f  \ \simeq\  [b,F] \quad,\quad
\pa_\zb f\ \simeq\ -[\bdag,F] \quad,& \\[1ex] &
\textrm{and}\qquad
\smallint\!\diff^4x\;f(x)\ =\ (2\pi\th)^2\,\tr_{\Hcal_2} F \quad, &
\eea
where the two-oscillator Fock space $\Hcal_2$ is spanned by
\be
|m_1,m_2\>\=\sfrac{1}{\sqrt{m_1! m_2!}}\,(\adag)^{m_1}\,(\bdag)^{m_2}\,|0,0\>
\qquad\textrm{with}\quad m_1,m_2\in\NN_0 \quad.
\ee

\section{D=2+1: Solitons in noncommutative Yang-Mills-Higgs}
\noindent
Starting from $D=2{+}2$ with signature $(++-\,-)\ $ I put $\th'=0$ and
\be
A_3=\p \quad,\qquad \pa_3=0 \quad,\qquad x^4=-t \quad,\qquad 
y\=x^1+\ii x^2 \=\sqrt{2\th}\,a\quad.
\ee
Instead of using the hermitean gauge (see Section~3) it is more
convenient in this situation to switch to the so-called unitary gauge.
For the spectral parameter this entails a transformation from the unit
disk to the upper half plane,
\be
\la\quad\to\quad\z \= \ii\,\sfrac{1+\la}{1-\la} 
\qquad\qquad\textrm{so that}\qquad 1/\lb\quad\to\quad\bar{\z} \quad.
\ee
This time, the two combinations $A_t{-}A_2$ and $A_1{+}\p$ are gauged away, 
but now staying within~U$(n)$ so that the normalization condition remains
\be \label{Res2}
A_\mu^\+ \= -A_\mu \qquad\Longleftrightarrow\qquad
\Ps(x,\z)\,\Ps(x,\bar{\z})^\+ \= \bsy{1} \quad.
\ee
The linear system looks slightly different, and $\Ps$ produces $A$ via
\be \label{Res1}
2 A_1 =\Ps(\z)\bigl(\pa_1 - \z(\pa_t{-}\pa_2)\bigr)\Ps(\z)^{-1}
\qquad\textrm{\&}\qquad
2 A_2 =\Ps(\z)\bigl((\pa_t{+}\pa_2) - \z\pa_1\bigr)\Ps(\z)^{-1} \quad.
\ee
Consequently, the single-pole ansatz has to be modified to
\be \label{ansatz}
\Ps(a,\adag,t,\z)\=\bsy{1}+\frac{\m-\bar{\m}}{\z-\m}\,P(a,\adag,t) \quad.
\ee
Again, the absence of poles at $\z{=}\m$ or $\z{=}\bar{\m}$ in
(\ref{Res2}) and (\ref{Res1}) leads to
\bea
(\ref{Res2})\qquad &\Longrightarrow &\qquad
P^2 \= P \= P^\+ 
\qquad \Longleftrightarrow \qquad
P\=T\,\sfrac{1}{\Tdag T}\,\Tdag
\\[8pt] \label{Eigen1}
(\ref{Res1})\qquad &\Longrightarrow &\qquad
(\bsy{1}{-}P)\,L\,P \= 0 
\qquad \Longleftrightarrow \qquad
L\,T\= T\,\g 
\eea
where $L$ denotes the differential operators in~(\ref{Res1})
for $\z{=}\bar{\m}$ and $\g$ is some $r{\times}r$ matrix.

In the noncommutative setup~\cite{ncsol} the $(x^1,x^2)$ coordinate dependence
gets traded for operator valuedness while the time~$t$ remains a parameter.
The Heisenberg algebra $[a,\adag]=1$, when represented on $\Hcal_1$ with 
basis $\{|m\>, m{\in}\NN_0\}$, turns $A$, $\Ps$ and $P$ into semi-infinite
$n\infty{\times}n\infty$ matrices acting on $\C^n\otimes\Hcal_1$ for the
gauge group~U($n$). The collection $T$ of $r$ column vectors is then seen
as an $n{\times}r$ array of $\Hcal_1$ kets:
\be
|T\> \= \Bigl( |T_i^{\ \ell}\> \Bigr)_{i=1\ldots n}^{\ell=1\ldots r} 
\qquad\Longrightarrow\qquad
P \= |T\>\,\sfrac{1}{\<T|T\>}\,\<T| \quad.
\ee

As detailed in~\cite{ncsol},
the time dependence of the two operators~$L$ in~(\ref{Res1})
can be absorbed into an ISU(1,1) coordinate transformation from $a$ to
\bea
&& c \= (\cosh\tau)\,a - (\e^{\ii\vartheta}\sinh\tau)\,\adag - \b\,t
\= U(t)\,a\,U^\+(t) \\[4pt]
&& \textrm{with}\qquad
U(t)\=\e^{\a\,\adag\adag-\bar{\a}\,a\,a}\;\e^{(\b\,\adag-\bar{\b}\,a)t}\quad,
\eea
where the parameters $\tau,\vartheta,\b,\a$ are specific functions 
of~$\m$ only. These ``moving-frame'' coordinates are canonical, 
$[c,\cdag]=1$, and describe moving squeezed states
\be
|m\>_t \= U(t)\,|m\> \qquad\textrm{based on}\quad c\,|0\>_t \=0 \quad.
\ee
It turns out that both operators~$L$ are proportional to $[c,.]$, 
so that (\ref{Eigen1}) mutates to
\be \label{Eigen2}
c\,|T_i^{\ \ell}\> \= 
|T_i^{\ \ell'}\>\,\g_{\ell'}^{\ \ell}
\qquad\textrm{for some $r{\times}r$ matrix}\quad 
\g = (\g_{\ell'}^{\ \ell}) \quad.
\ee
This equation may be interpreted as a holomorphy condition.
It is solved by any collection of states which spans a subspace
of $\C^n\otimes\Hcal_1$ invariant under the action of~$c$. 
Each solution yields a time-dependent exact noncommutative U($n$) soliton
\be
2 A_1 \= (\m{-}\mb)\,(\pa_t-\pa_2)\,P \qquad,\quad
2 A_2 \= (\m{-}\mb)\, \pa_x\,P
\ee
with a topological charge~$q$ and energy 
$E=8\pi q \frac{\sqrt{1-v_1^2-v_2^2}}{1-v_2^2}$.
Its energy density consists of up to $q$ lumps moving jointly in the
$(x^1,x^2)$ plane with a constant velocity $(v_1,v_2)$ given by~$\m$.
Having zero {\it relative\/} lump velocities these configurations
cannot be considered as true {\it multi-\/}solitons.
For $\ \m=-\ii\ $ one finds $\vec{v}=0$ and $c=a$, 
i.e. the static case $\ U(t)=1$.

Let us first look at solutions with finite rank~$r<\infty$.
An interesting class of solutions occurs for diagonal matrices
$\ \g=\textrm{diag}(\g_1,\ldots,\g_r)$, because then (\ref{Eigen2}) 
decouples to
\be
c\,|T_i^{\ \ell}\> \= |T_i^{\ \ell}\>\,\g_\ell 
\qquad\forall\ i=1,\ldots,n \quad\textrm{and}\quad \ell=1,\ldots,r 
\ee
which is solved by coherent states based on the squeezed vacuum, 
as illustrated for $n{=}1$:
\be
|T^\ell\>\ \sim\ \e^{\g_\ell\cdag}|0\>_t \= U(t)\,\e^{\g_\ell\adag}|0\>
\ \sim\ U(t)\,|\g_\ell\> \= |\g_\ell\>_t \quad. 
\ee
Obviously, $\g_\ell$ simply gives the position of the $\ell$th lump 
at $t{=}0$. For the simplest case, $r{=}1$ and $\vec{v}{=}0$, one finds
\be
P \= |\g\>\<\g| \= 
\textrm{Weyl-order} \bigl[ 2\,\e^{-2(\adag-\bar{\g})(a-\g)} \bigr]
\qquad\Longrightarrow\qquad 
p(y) \= 2\,\e^{-|y-\sqrt{2\th}\g|^2/\th}
\ee
which becomes singular in the commutative limit $\th{\to}0$.
Any (abelian as well as non-abelian) finite-rank solution~$|T\>$ 
has topological charge $q=r$ and can be built from such coherent states. 
Moreover, $|T\>$ is always unitarily related to the standard choice 
$(|0\>,|1\>,\ldots|r{-}1\>)$ belonging to $\ (\g)^r=0$. In this sense,
all finite-rank solutions have abelian character, as exemplified here
for $n{=}2$, $r{=}2$ at $\vec{v}{=}0$:
\be
\left(\begin{matrix} |0\> & |1\> \\[6pt] 0 & 0 \end{matrix}\right) \=
\left(\begin{matrix} |0\>\<0| & S^\+ \\[6pt] S & 0 \end{matrix}\right)
\left(\begin{matrix} |0\> & 0 \\[6pt] 0 & |0\> \end{matrix}\right) \quad,
\ee
where $S=\frac{1}{\sqrt{a\adag}}a\,:\ |m\>\to|m{-}1\>$ is the shift operator.

Infinite-rank solutions can also feature {\it finite\/} topological charge if
they are truely nonabelian. As a static U(2) example with $\ r=1{+}\infty\ $
but $\ q=1\ $ consider
\be
|T\> \= \left(\begin{matrix} |0\>\<0| \\[6pt] S \end{matrix}\right) 
\bigl( |0\>\ |1\>\ |2\>\ |3\>\ \ldots \bigr) \=
\left(\begin{matrix} |0\> & 0 & 0 & 0 & \ldots \\[6pt] 
                     0 & |0\> & |1\> & |2\> & \ldots \end{matrix}\right)  
\ee
which yields $\ P=|0\>\<0|\oplus1\ $ on $\ \Hcal_1\oplus\Hcal_1$.
Using $\ S|0\>=0\ $ and $\ SS^\+=1\ $ but $\ S^\+S=1{-}|0\>\<0|$, 
the unitary transformation
\be
\Biggl(\begin{matrix} 
\sfrac{\eps}{K}|0\>\<0|-\sfrac{\sqrt{\adag a}}{K} &\sfrac{\eps}{K}\,S^\+\\[8pt]
S\,\sfrac{\bar\eps}{K} & S\sfrac{\sqrt{\adag a}}{K} S^\+ \end{matrix}\Biggr)
\left(\begin{matrix} |0\>\<0| \\[10pt] S \end{matrix}\right) \=
\left(\begin{matrix} \eps \\[10pt] a \end{matrix}\right) \, \frac{1}{K}
\ee
with $\ K=\sqrt{\adag a+\bar\eps\eps}\ $ and parametrized by $\eps\in\C$
clearly maps
\be
|T\>\ \to\ |T(\eps)\> \= |\Tcal\> \sfrac{1}{\sqrt{\<\Tcal|\Tcal\>}}
\qquad\textrm{with}\quad 
|\Tcal\> \= \Bigl(\begin{matrix} \eps \\ a \end{matrix}\Bigr)
\bigl( |0\>\ |1\>\ |2\>\ \ldots \bigr) \quad.
\ee
Since $\ \lim_{\eps\to0} |T(\eps)\> = |T\>\ $ the parameter~$\eps$ may be
regarded as a regulator.
The complete basis of kets drops out when building the projector,
and so the choice of~$|\Tcal\>$ is equivalent to taking 
\be
T \= \Tch\ \equiv\ \biggl(\begin{matrix} \eps \\ a \end{matrix}\biggr)
\qquad\Longrightarrow\qquad P \= 
\Biggl(\begin{matrix} \sfrac{\eps\bar\eps}{K} & \sfrac{\eps}{K}\adag \\[8pt]
a \sfrac{\bar\eps}{K} & a\sfrac{1}{K}\adag \end{matrix}\Biggr) \quad.
\ee
Quite generally, for $\ |T\>=\Th\,\bigl(|0\>\ |1\>\ |2\>\ \ldots \bigr)\ $
with $\Th$ being an $n{\times}r'$ array of operators,
the condition (\ref{Eigen2}) translates to
\be
[\,c\,,\Th\,] \= \Th\ \gh \qquad
\textrm{for some $r'{\times}r'$ array}\ \gh\ \textrm{of operators}\quad.
\ee
In case $\ \gh=\textrm{diag}(\g_1,\ldots,\g_{r'})\ $ with c-numbers $\g_\ell$
the solution takes the simple form
\be
\Th_i^{\ \ell} \= \e^{\g_\ell \cdag}\,f_i^{\ \ell}(c) 
\ee
with arbitrary functions $f_i^{\ \ell}$ of $c$ only.
The corresponding projectors have infinite rank in $\C^n\otimes\Hcal_1$
but the topological charge (and the energy) is determined by the degrees of
$f_i^{\ \ell}$ if the latter are polynomial.

These nonabelian solutions possess a regular $\th\to0$ limit 
where they coincide with the known commutative solutions, 
which live in the Grassmannian 
$\ {\rm Gr}(n,r')=\frac{{\rm U}(n)}{{\rm U}(r')\times{\rm U}(n{-}r')}$.
Formally, their noncommutative deformations are elements of
$\ {\rm Gr}(n\infty,r{=}r'\infty{+}q)$, 
and the abelian solitons are included at $(n{=}1,r'{=}0)$.

Proper multi-solitons~\cite{ncsol} can be constructed by iterated dressing 
of (\ref{ansatz}):
\be \label{multi}
\Ps_m(a,\adag,t,\z) \= \prod_{k=1}^m
\Bigl( \bsy{1}-\frac{\m_k-\bar{\m}_k}{\z-\m_k}\,P_k(a,\adag,t) \Bigr) \=
\bsy{1} - \sum_{k=1}^m \frac{R_k(a,\adag,t)}{\z-\m_k}
\ee
if all $\m_k$ are mutually different.
Repeating the previous analysis of the pole structure yields $P_k$ and
$R_k$ in terms of $m$ moduli $\m_k$ and ket matrices $|T\>_k$ subject to
\be
(\bsy{1}_n\otimes c_k)|T\>_k \= |T\>_k\cdot\g_k
\qquad\textrm{where}\qquad c_k\=U_k(t)\,a\,U_k^\+(t)
\ee		      
is the ``moving-frame'' coordinate derived from $\m_k$.
In this fashion one arrives at $m$ copies of the single-pole solution;
the corresponding (clusters of) lumps, however, move at mutually different 
velocities $\vec{v}_k$! 
Their topological charges and energies are simply additive.
I should stress that the time dependence of these configurations is 
{\it exact\/} and not just valid in the adiabatic regime.

In this situation the question of scattering immediately emerges.
An analysis of the asymptotic behavior shows that the ansatz (\ref{multi})
can only lead to no-scattering solutions, i.e. the lumps do not disturb
one another. However, a slight generalization allowing for coinciding poles
in (\ref{multi}) produces nonabelian multi-solitons which scatter at angles 
$\ \vt=\pi/\ell\ $ as well as abelian breather-type configurations~\cite{scat}.

\section{D=4+0: Instantons in noncommutative Yang-Mills} 
\noindent
I will construct a self-dual finite-action U(2) configuration 
on a self-dual noncommutativity background. Not fixing the gauge,
the basic relations are (\ref{AfromPsi}) and (\ref{norm}).
A one-instanton ansatz for $\Ps$ with poles at $\la{=}0$ and $\la{=}\infty$
reads~\cite{instcom,hlw}
\be 
\Ps(x,\la) \= G(x)\,\bigl(H(x)+\la S(x)^\+ +\sfrac{1}{\la}S(x) \bigr)\quad.
\ee
On this I impose the following restrictions:
\be
G\=\bigl(\begin{smallmatrix} g_- & 0 \\ 0 & g_+ \end{smallmatrix}\bigr)\=G^\+
\quad,\quad
H\=\bigl(\begin{smallmatrix} h_- & 0 \\ 0 & h_+ \end{smallmatrix}\bigr)\=H^\+
\quad,
\ee 
\be \label{comm}
[G,H]\=[G,S]\=[H,S]\=0\quad.\phantom{XXX}
\ee
Inserting the above into (\ref{AfromPsi}) and (\ref{norm})
and separating different powers of~$\la$ produces
\bea
(\ref{norm}) \quad\Longrightarrow && 
S^2\=0\ ,\quad G^2 H^2 \= 1+G^2\{S,S^\+\} \quad, \label{SS} \\[12pt]
(\ref{AfromPsi}) \quad\Longrightarrow && 
S\pa_\yb S\=0\=S\pa_\zb S \quad, \\[4pt] && 
H\pa_\yb S-S\pa_\yb H-S\pa_z S\=0 \quad, \label{HS1} \\[4pt] && 
H\pa_\zb S-S\pa_\zb H+S\pa_y S\=0 \quad. \label{HS2}
\eea

A convenient choice for $S$ is
\be
S\=\biggl(\begin{smallmatrix} 
z\frac{1}{f(r)}y\ & z\frac{-1}{f(r)}z \\[4pt] 
y\frac{1}{f(r)}y\ & y\frac{-1}{f(r)}z
\end{smallmatrix}\biggr)
\qquad\textrm{with}\qquad 
r^2\ :=\ \yb y+\zb z \= 2\th\,(\adag a+b\,\bdag) \quad.
\ee
Then (\ref{comm}) is solved by
$\ g_\pm(x)=g(r^2\pm2\th)\ $ and $\ h_\pm(x)=h(r^2\pm2\th)$, 
and furthermore
\bea
(\ref{HS1},\ref{HS2})\quad\Longleftarrow && 
h(r^2)\=-1 \qquad\textrm{and}\qquad f(r)\=r^2+2\Lambda^2 \quad,\\[4pt]
(\ref{SS})  \quad\Longrightarrow && 
g(r^2)\=\pm\frac{1}{2\Lambda}\,\frac{r^2+2\Lambda^2}{\sqrt{r^2+\Lambda^2}}
\quad,
\eea
with some real parameter $\Lambda$.

Putting it all together I arrive at
\be
\Ps \= \frac{1}{2\Lambda}
\begin{pmatrix} \frac{1}{\sqrt{r^2+\Lambda^2-2\th}} & 0 \\[16pt]
                0 & \frac{1}{\sqrt{r^2+\Lambda^2+2\th}} 
\end{pmatrix} 
\begin{pmatrix} {\scriptstyle r^2+2\Lambda^2-2\th-\la\yb\zb-}\frac{yz}{\la}
              & {\scriptstyle -\la\yb^2+}\frac{z^2}{\la} \\[16pt] 
                {\scriptstyle  \la\zb^2-}\frac{y^2}{\la}
              & {\scriptstyle r^2+2\Lambda^2+2\th+\la\yb\zb+}\frac{yz}{\la}
\end{pmatrix}
\ee
which finally yields the noncommutative BPST instanton 
(see also~\cite{furuuchi}):
\vspace{2mm}
\bea
{A}_{y} &=& \begin{pmatrix}
-\frac{\bar{y}}{2\th}
\Bigl(\sqrt{\frac{r^2+\Lambda^2-2\th}{r^2+\Lambda^2}} - 1 \Bigr) & 
0 \\[14pt]
-\bar{z}\,\frac{1}{\sqrt{r^2+\Lambda^2}\sqrt{r^2+\Lambda^2-2\th}} &
-\frac{\bar{y}}{2\th}
\Bigl(\sqrt{\frac{r^2+\Lambda^2+4\th}{r^2+\Lambda^2+2\th}} - 1 \Bigr) 
\end{pmatrix} \quad, \\[20pt]
{A}_{z} &=& \begin{pmatrix}
\Bigl(\sqrt{\frac{r^2+\Lambda^2-2\th}{r^2+\Lambda^2}} - 1 \Bigr)
\frac{\bar{z}}{2\th} \quad &
-\frac{1}{\sqrt{r^2+\Lambda^2}\sqrt{r^2+\Lambda^2-2\th}}\,\bar{y}
\ \ \,\\[14pt] 0 & 
\Bigl(\sqrt{\frac{r^2+\Lambda^2+4\th}{r^2+\Lambda^2 +2\th}} - 1 \Bigr)
\frac{\bar{z}}{2\th} 
\end{pmatrix} \quad.
\eea

\section{D=3+0: Monopoles in noncommutative Yang-Mills-Higgs}
\noindent
Beginning from $D=4{+}0\ $ I set $\ \th'=0\ $ and
\be
A_4=\p \quad\textrm{and}\quad \pa_4=0 \qquad\Leftrightarrow\qquad
\pa_z-\pa_\zb=0 \quad\textrm{and}\quad  D\p=*F
\ee
One may again try to apply the dressing method, but it turns out that
this situation is more amenable to the (related) splitting method,
which reformulates the linear system~(\ref{linsys}) as a parametric
Riemann-Hilbert problem~\cite{splitting}. 
Lacking the time to explain this approach in any detail I will 
only sketch the salient features for the monopole case~\cite{ncmon}.

In all methods, the BPS equation gets reduced to a linear (differential)
equation for an auxiliary object, 
e.g. for $T$ in (\ref{Eigen1}) or (\ref{Eigen2}), which restricts the
coordinate dependence of this quantity. Within the splitting approach, 
it is a matrix-valued so-called transition function~$f_{+-}(x,\la)$ which in
$D{=}3$ depends on the coordinates~$\ x=(x^1,x^2,x^3)\ $ only holomorphically
through the combination (remember $\ y=x^1+\ii x^2\ $ and $\ [y,\yb]=1$)
\be 
w(\la)\=2x^3+\la\yb-\la^{-1}y \quad.
\ee
The Riemann-Hilbert task of factorizing 
\be
f_{+-}(x,\la)\= \Ps_+^{-1}(x,x^4,\la)\,\Ps_-(x,x^4,\la)
\qquad\textrm{with}\quad \Ps_\pm(\la)\ \textrm{holomorphic for\ } 
|\la|\,\begin{smallmatrix} < \\ > \end{smallmatrix}\,1
\ee 
requires the multiplicative and additive decompositions
\be
w \= \sfrac{1}{\la}\,(y\x^{-1}+\la\x)\,(\la\x^{-1}\yb-\x)
  \= (z+\la\yb)\,-\, (\la^{-1}y-\zb) \= u\,-\,v \quad,
\ee
where $\x$ was computed in~\cite{gn} 
and obeys $\ \x^2\com r-x^3\ $ with $\ r^2=\yb y+x^3 x^3$.

In case of $u(2)$ monopoles, the simplest ansatz for the $2{\times}2$ 
matrix~$f_{+-}$ contains a function~$\r$ which for the BPS monopole 
can be reduced to the Weyl-ordered expression
\bea
\r &=& \e^{u}\,w^{-1}\,\e^{u}-\e^{v}\,w^{-1}\,\e^{v} \\
   &=& \e^{-2\ii x^4}\,\smallint_{-1}^{+1}\!\!\diff{t}\;\e^{2t x^3}\,
       \e^{\la(1+t)\yb+\la^{-1}(1-t)y} \\
   &=& \e^{-2\ii x^4} \,\textstyle{\sum_{m\in\Z}} \,\r_m\,\la^m \\[2ex] 
\textrm{where}\qquad
\r_0 &=& \sinh(2R)/R \qquad\textrm{with}\quad R=x^3+\x\x\ \com\ r\quad,\\[4pt]
\r_{\pm1} &=& \textrm{explicitly known in terms of ($y,\yb,x^3,R$)} \quad.
\eea 
The gauge potential is entirely expressed in terms of~$\r_0$ via~\cite{ncmon}
\bea 
A_i \= \;\ve_{ijk}\,\frac{\s_k}{2\ii}\!\!\!\!
&\Bigl(\r_0^{+\frac12}\pa_j\,\r_0^{-\frac12}-
\r_0^{-\frac12}\pa_j\,\r_0^{+\frac12}\Bigr)& \\[1ex]
+\ \;\frac{\bsy{1}_2}{2}\!\!\!\!
&\Bigl(\r_0^{-\frac12}\pa_i\,\r_0^{+\frac12}+
\r_0^{+\frac12}\pa_i\,\r_0^{-\frac12}\Bigr)&\!\!\! 
+\ \;\s_i \quad,\\[1.5ex] 
\p\ \equiv\ A_4\ =\ \;\frac{\s_i}{2\ii}\!\!\!\!
&\Bigl(\r_0^{+\frac12}\pa_i\,\r_0^{-\frac12}-
\r_0^{-\frac12}\pa_i\,\r_0^{+\frac12}\Bigr)& \ .
\eea
The expression for~$A_i$ is not antihermitean and was not expected to be
because I did not impose a reality condition on the factorization on~$f_{+-}$.
However, it is possible to pass to an antihermitean configuration
via a nonunitary gauge transformation generated by
\be
g\=\bigl[\Ps_+(\la)\,\Ps_-^{\dagger}(-1/\lb)\bigm|_{\la=0}\bigr]^{1/2}\quad.
\ee
Unfortunately, the matrix $g^2$ is quite complicated and involves 
$\r_{\pm1}$ as well~\cite{ncmon}. Reassuringly, the commutative limit 
reproduces the familiar result:
\be
A_i \= \ve_{ijk}\,\frac{\s_k}{2\ii}\,\frac{x_j}{r}
\Bigl(\frac{1}{r}-2\,{\rm coth}(2r)\!\Bigr)\,+\;\s_i
\qquad\buildrel{g=\e^{x^i\s_i}}\over{\hbox to 40pt{\rightarrowfill}}\qquad
\ve_{ijk}\,\frac{\s_k}{2\ii}\,\frac{x_j}{r}
\Bigl(\frac{1}{r}-\frac{2}{\sinh(2r)}\Bigr) \quad.
\ee

\section{Other applications}

The methods outlined in this talk have also been applied successfully towards 
the construction and study of various other noncommutative field configurations
(see for example \cite{hoissen,hamanaka,2d,wadati} and references 
therein\footnote{
This is not a review. I apologize for my incomplete citation.}). 
In my group, in particular, we have investigated
\cite{n2,swmon,duy,n2def,ssft}
\begin{itemize}
\item
Relations of noncommutative integrable models with open $N{=}2$ strings
\item
Seiberg-Witten monopole equations on $\R^4_\th$, related with
vortex-type equations on $\R^2_\th$
\item
Donaldson-Uhlenbeck-Yau equations on $\ \R^{2n}_\th\times S^2\ $
give vortex-type equations on $\R^{2n}_\th$
\item
Moyal-deformed extended superspace and gauge theory thereon
\item
Open superstring field theory (\`a la Witten or Berkovits),
which can be interpreted as an integrable infinite-dimensional
noncommutative field theory
\end{itemize}
Numerous open problems remain to be tackled, such as
finding nontrivial classical superstring configurations,
analyzing the quantum fluctuations around our BPS solutions,
or substantiate their D-brane interpretation.


\bigskip\noindent
{\bf Acknowledgement} 

\noindent
I am grateful to my coworkers who helped bring this program to life.


\end{document}